\documentclass[12pt]{iopart}   
\usepackage{iopams}  
\usepackage{color} 
\usepackage[sort]{cite}
\usepackage{graphicx}  
\usepackage{dcolumn}   
\usepackage{bm}        
\usepackage{amssymb}   
\usepackage[english]{babel}

\begin{document}

\title{Magnetic-field-dependent photodynamics of single NV defects in diamond: Application to qualitative all-optical magnetic imaging}
\date{\today}

\author{J-P Tetienne$^1$, L Rondin$^1$, P Spinicelli$^1$, M Chipaux$^{2}$, T~Debuisschert$^2$, J-F Roch$^{1,3}$ and V Jacques$^1$}

\address{$^{1}$Laboratoire de Photonique Quantique et Mol\'eculaire, CNRS and ENS Cachan UMR 8537, 94235 Cachan, France}
\address{$^{2}$Thales Research \& Technology, Campus Polytechnique, 91767 Palaiseau, France}
\address{$^{3}$Laboratoire Aim\'e Cotton, CNRS, Universit\'e Paris-Sud and ENS Cachan, 91405 Orsay, France}
\ead{vjacques@lpqm.ens-cachan.fr}

\begin{abstract}
\vspace{0.3cm}

\noindent Magnetometry and magnetic imaging with nitrogen-vacancy (NV) defects in diamond rely on the optical detection of electron spin resonance (ESR).  However, this technique is inherently limited to magnetic fields that are weak enough to avoid electron spin mixing. Here we focus on the high off-axis magnetic field regime for which spin mixing alters the NV defect spin dynamics. We first study in a quantitative manner the dependence of the NV defect optical properties on the magnetic field vector {\bf B}. Magnetic-field-dependent time-resolved photoluminescence (PL)  measurements are compared to a seven-level model of the NV defect  that accounts for field-induced spin mixing. The model reproduces the decreases in (i) ESR contrast, (ii)  PL intensity and (iii) excited level lifetime with an increasing off-axis magnetic field. We next demonstrate that those effects can be used to perform all-optical imaging of the magnetic field component $|B_{\perp}|$ orthogonal to the NV defect axis. Using a scanning NV defect microscope, we map the stray field of a magnetic hard disk through both PL and fluorescence lifetime imaging. This all-optical method for high magnetic field imaging at the nanoscale might be of interest in the field of nanomagnetism, where samples producing fields in excess of several tens of milliteslas are typically found.
\end{abstract}

\maketitle 
 
\section{Introduction} 

The nitrogen-vacancy (NV) defect in diamond is a solid-state quantum system that has been extensively studied over the last years \cite{Jelezko2006}. Potential applications include quantum information processing \cite{Dutt2007,Neumann2010,Robledo2011b,Fuchs2011}, imaging in life science~\cite{Liam2011}, hybrid quantum systems \cite{Kubo2011,Arcizet2011,Shimon2012} and magnetic sensing and imaging \cite{Degen2008,Taylor2008,Maze2008,Balasubramanian2008}. For the latter application, the magnetic field is evaluated by measuring the Zeeman shifts of the NV defect electron spin sublevels through the optical detection of electron spin resonance (ESR). The main advantage of NV-based magnetometry is the possible combination of atomic-scale spatial resolution with high magnetic field sensitivity -- below 10 nT/Hz$^{1/2}$~\cite{Balasubramanian2009} -- even under ambient conditions. Magnetic imaging with diffraction-limited micrometer resolution has been demonstrated using an ensemble of NV defects~\cite{Maertz2010,Steinert2010,Pham2011} while recent experiments have reached the nanoscale with a single NV defect coupled to a scanning probe microscope~\cite{Balasubramanian2008,Maletinsky2012,Rondin2012}. 
 
However, ESR-based magnetometry with NV defects is intrinsically limited to magnetic fields with an amplitude and an orientation such that the electron spin quantization axis remains fixed by the NV defect axis itself. Indeed, any significant spin mixing induced by an off-axis magnetic field rapidly reduces the contrast of optically-detected ESR spectra because optically induced spin polarization and spin dependent photoluminescence (PL) of the NV defect become inefficient. Besides a decreased ESR contrast, the PL intensity as well as the effective excited level lifetime are observed to decrease with an increasing off-axis magnetic field~\cite{Epstein2005,Diep2009}. It was recently shown that this property can be used as a resource to perform all-optical magnetic field mapping with a scanning NV defect~\cite{Maletinsky2012,Rondin2012}. The purpose of the present work is to study in a quantitative manner the dependence of the NV defect optical properties on the magnetic field vector {\bf B}, and to discuss the implications for magnetic field imaging. The paper is organized as follows. In Section~\ref{sec1}, we first investigate the dynamics of optically-pumped NV defects as a function of {\bf B}, focusing on (i) the ESR contrast, (ii) the PL intensity and (iii) the excited level lifetime. A simple seven-level model of the NV defect is developed and compared to magnetic-field-dependent time-resolved PL measurements. This method provides a simple way to extract all relevant photophysical parameters of the NV defect and compute its optical response as a function of the $\mathbf{B}$ field. The results of the calculation are then compared to single site measurements and discussed in the context of magnetic field imaging. In Section~\ref{sec2}, we report on experiments of all-optical magnetic imaging in the high off-axis magnetic field regime using a scanning NV defect microscope. In particular, fluorescence lifetime imaging (FLIM) capabilities allows us to record not only PL images but also lifetime ones, a feature that could reveal useful beyond magnetic imaging, {\it e.g.} for mapping the local density of electromagnetic states (LDOS) of photonic nanostructures~\cite{Farahani2005,Frimmer2011}.

\section{Photodynamics of single NV defects in a static magnetic field} \label{sec1}

\subsection{Model}
\begin{figure}[t]
\begin{center}
   \includegraphics[width=0.9\textwidth]{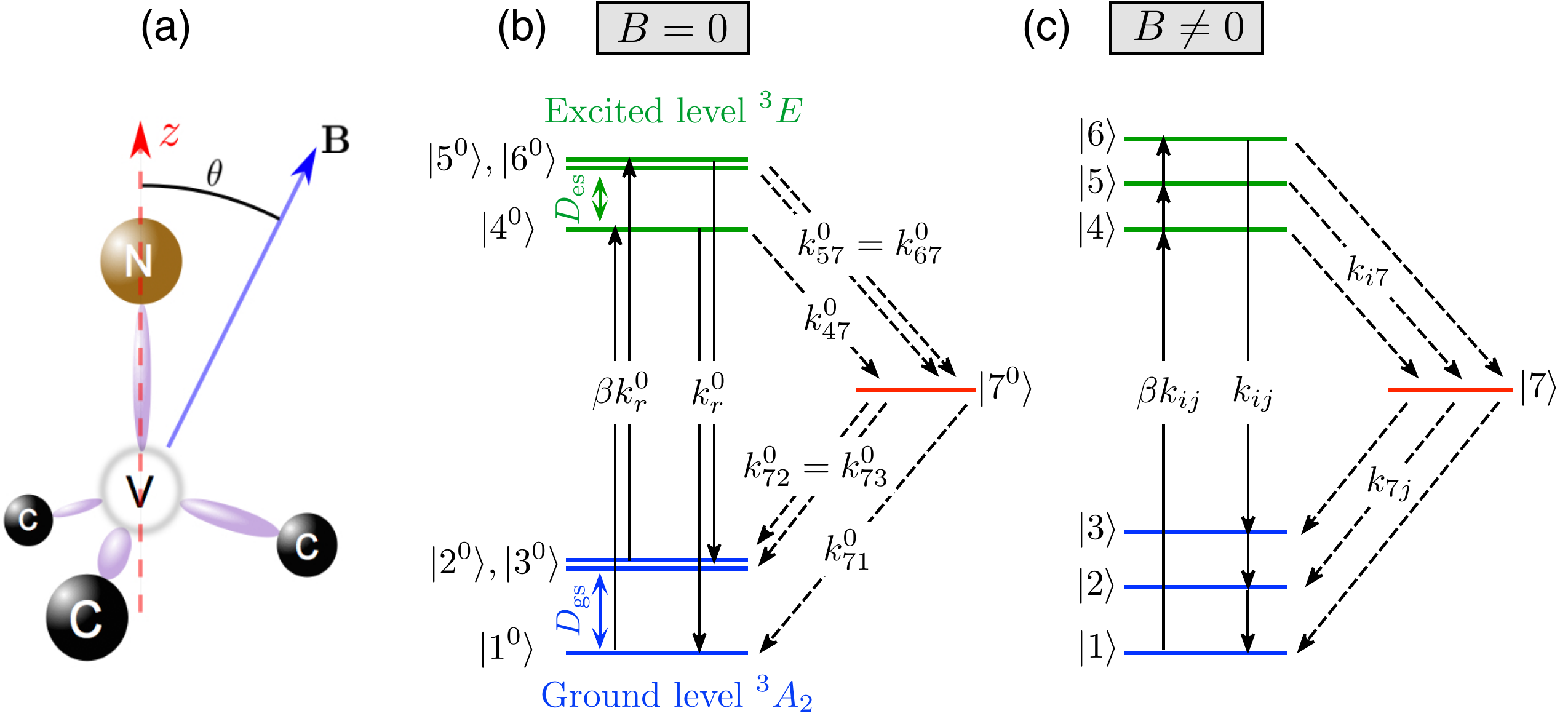}
\caption{(a) Atomic structure of the NV defect in diamond. A magnetic field $\mathbf{B}$ is applied with an angle $\theta$ with respect to the NV defect axis $z$. (b,c) Seven-level energy level structure considered for the NV defect at room temperature. The ground ($^3A_2$) and excited ($^3E$) spin triplets are linked by spin conserving radiative transitions (solid lines) while the singlet states, merged into state $\vert7\rangle$, provide a non-radiative relaxation path (dashed lines). All notations are defined in the main text. (b) At $B=0$, five intrinsic parameters $\lbrace k_{r}^0,k_{47}^0,k_{57}^0,k_{71}^0,k_{72}^0\rbrace$ are used to describe the NV defect photodynamics. (c) For a given applied magnetic field $B$, all transition rates $k_{ij}$ are likely to be non vanishing owing to electron spin mixing both in the ground level and in the excited level.}
\label{Fig1}
\end{center}
\end{figure}

\indent The negatively charged NV defect in diamond consists of a substitutional nitrogen atom (N) associated with a vacancy (V) in an adjacent lattice site of the diamond matrix, giving a defect with $C_{\rm 3v}$ symmetry~(Fig.~\ref{Fig1}(a)). Its ground level is a spin triplet $^{3}A_{2}$, whose degeneracy is lifted by spin-spin interaction into a singlet state of spin projection $m_{s}=0$ and a doublet $m_{s}=\pm 1$, separated by $D_{\rm gs}=2.87$~GHz in the absence of magnetic field~\cite{Manson2006}. Here, $m_{s}$ denotes the spin projection along the NV defect axis $z$~(Fig.~\ref{Fig1}(a)). When a magnetic field $\mathbf{B}$ is applied to the defect, the ground level spin Hamiltonian reads 
\begin{equation}
\label{Hamilto}
{\cal H}_{\rm gs}=h D_{\rm gs}S_{z}^{2}+g\mu_B\mathbf{B}\cdot\mathbf{S}={\cal H}_{\rm gs}^z+{\cal H}_{\rm gs}^\perp
\label{Eq1}
\end{equation}
with ${\cal H}_{\rm gs}^z=h D_{\rm gs}S_{z}^{2}+g\mu_B B_z S_z$ and ${\cal H}_{\rm gs}^\perp=g\mu_B(B_x S_x+B_y S_y)$, where $h$ is the Planck constant, $\mu_B$ is the Bohr magneton, and $g\approx 2$ is the electron $g$-factor. We note that strain-induced splitting and hyperfine interaction with nearby nuclear spins are neglected in Eq.~(\ref{Hamilto}). In zero field, the quantization axis is thus fixed by the NV defect axis and the eigenstates are labelled $\vert1^0\rangle$, $\vert2^0\rangle$ and $\vert3^0\rangle$, corresponding to spin projections $m_s=0$, $m_s=-1$ and $m_s=+1$, respectively (Fig.~\ref{Fig1}(b)). 

\indent The defect can be optically excited through a dipole-allowed transition to a $^{3}E$ excited level, which is also a spin triplet. Besides, the $^{3}E$ excited level is an orbital doublet which is averaged at room temperature~\cite{Rogers2009,Batalov_PRL} leading to a zero-field splitting $D_{\rm es}=1.42$~GHz with the same quantization axis and similar gyromagnetic ratio as in the ground level~\cite{Fuchs2008,Neumann2009}. The excited level spin Hamiltonian ${\cal H}_{\rm es}$ is thus simply given by Eq.~(\ref{Hamilto}) while replacing $D_{\rm gs}$ by $D_{\rm es}$. The excited level eigenstates in zero field are labelled $\vert4^0\rangle$, $\vert5^0\rangle$ and $\vert6^0\rangle$, corresponding to spin projections $m_s=0$, $m_s=-1$ and $m_s=+1$, respectively (Fig.~\ref{Fig1}(b)). Once optically excited in the $^{3}E$ level, the NV defect can relax either through the same radiative transition producing a broadband red photoluminescence (PL), or through a secondary path involving non radiative intersystem crossing (ISC) to singlet states. Recent experiments have identified two singlet states ($^{1}E$,$^{1}A_{1}$)~\cite{Rogers_2008,Acosta}, whereas theoretical studies predict the presence of a third singlet state between the ground and excited triplet levels~\cite{{Ma2010}}. In the present work, the singlet states are summarized into a single `metastable' level labelled $\vert7\rangle$. Spin-dependent ISC from the $^{3}E$ level to the `metastable' level is responsible for efficient spin-polarization in the $m_s=0$ spin sublevel through optical pumping as well as for spin-dependent PL of the NV defect. These two properties enable the detection of electron spin resonance (ESR) on a single NV defect by optical means~\cite{Gruber}.

 \indent Within this seven-level model of the NV defect, the zero-field transition rate from $\vert i^{0}\rangle$ to $\vert j^{0}\rangle$ is denoted by $k_{ij}^{0}$. We consider that optical transitions are purely spin-conserving, and that the radiative relaxation rate is spin-independent, {\it i.e.} $k_{41}^0=k_{52}^0=k_{63}^0=k_{r}^0$. This is reasonable since Robledo {\it et al.} recently determined that spin-flip radiative transition rates are at most $2\%$ of their spin-conserving counterpart \cite{Robledo2011}. On the other hand, the optical pumping rates from the ground to the excited level are proportional to the corresponding relaxation rates through $k_{ji}^{0}=\beta k_{ij}^{0}$ for $j=1,2,3$ and $i=4,5,6$, where $\beta$ is the optical pumping parameter. Lastly, ISC transition rates at zero-field only depend on the absolute value of $m_s$, {\it i.e.} $k_{57}^0=k_{67}^0$ and $k_{72}^0=k_{73}^0$. Following these assumptions, there are five intrinsic parameters left in the system, namely $k_{r}^0$, $k_{47}^0$, $k_{57}^0$, $k_{71}^0$ and $k_{72}^0$, and an extrinsic parameter $\beta$ linked to the optical pumping power (Fig.~\ref{Fig1}(b)).

\indent When a static magnetic field $\mathbf{B}$ is applied to the NV defect~(Fig.~\ref{Fig1}(c)), the seven eigenstates $\lbrace\vert i \rangle\rbrace$ of the system can be expressed as linear combinations of the zero-field eigenstates 
\begin{equation}
\label{eigen}
\vert i \rangle=\sum_{j=1}^7 \alpha_{ij}(\mathbf{B}) \vert j^0 \rangle \ ,
\end{equation}
where the coefficients $\lbrace\alpha_{ij}(\mathbf{B})\rbrace$ are numerically computed by using the expressions of $\mathcal{H}_{\rm gs}$ and $\mathcal{H}_{\rm es}$, and given that $\vert7\rangle=\vert7^{0}\rangle$. The new transition rates $\lbrace k_{ij}(\mathbf{B}) \rbrace$ are then related to the zero-field transition rates $\lbrace k_{ij}^0 \rbrace$ through the transformation
\begin{equation} \label{eqn3}
k_{ij}(\mathbf{B})=\sum_{p=1}^7 \sum_{q=1}^7 |\alpha_{ip}|^2 |\alpha_{jq}|^2 k_{pq}^0 \ ,
\end{equation}
which represents a statistical averaging using the weights of the initial and the final state of the transition. The lifetime of each eigenstate in the excited level is then given by $\tau_i=1/(\sum_{j=1}^7 k_{ij})$ with $i=4,5,6$.  

\indent With the aim of studying the time-resolved optical response of a single NV defect placed in a static magnetic field, we consider a pulsed optical excitation with a pulse duration $\delta t$ and a repetition period $T$ such that $T\gg\tau_i\gg\delta t$. After optical excitation ($t=0$), each excited state population $n_i$ decays exponentially as $n_i(t)=n_i(0){\rm e}^{-t/\tau_i}$ once averaged over many excitation cycles. The rate of emitted photons $\mathcal{R}_{i}(t,\mathbf{B})$ from state $\vert i \rangle$ is the radiative part of the total decay and reads $\mathcal{R}_{i}(t,\mathbf{B})=\sum_{j=1}^3 k_{ij} n_i(t)$. The detected time-resolved PL signal $\mathcal{R}(t,\mathbf{B})$ is thus finally obtained by summing over the three excited states 
 \begin{eqnarray} \label{eqn8}
\mathcal{R}(t,\mathbf{B})=\eta \sum_{i=4}^6 \sum_{j=1}^3 k_{ij} n_i(0) {\rm e}^{-t/\tau_i} ,
\end{eqnarray}  
where $\eta$ is the collection efficiency. In this equation, the magnetic field dependent transition rates $\lbrace k_{ij} \rbrace$ are given by Eq.~(\ref{eqn3}) and the set of coefficients $\lbrace n_i(0) \rbrace$ are inferred by solving the classical rate equations of the system
\begin{equation}
\label{rate}
\frac{{\rm d}n_i}{{\rm d}t}  =   \sum_{j=1}^7 (k_{ji}n_j - k_{ij}n_i) \ ,
\end{equation} 
while considering a closed seven-level model, {\it i.e.} $\sum_{i=1}^7 n_i=1$, and the periodic condition $n_i(t)=n_i(t+T)$. The detected time-resolved PL signal $\mathcal{R}(t,\mathbf{B})$ is therefore given by a tri-exponential decay, where the amplitude and the lifetime of each component is expressed as a function of the zero-field transition rates $\lbrace k_{r}^0,k_{47}^0,k_{57}^0,k_{71}^0,k_{72}^0\rbrace$ and the optical pumping parameter $\beta$. We expect that $n_4\gg (n_5,n_{6})$  at low magnetic field owing to efficient spin polarization of the NV defect in $m_s=0$ by optical pumping, while those populations are balanced in the limit of strong transverse magnetic fields.

Finally, we infer the contrast $\mathcal{C}$ of optically-detected ESR spectra for a single NV defect placed in a static magnetic field. For that purpose, non-vanishing transition rates $k_{12}=k_{21}$ (resp. $k_{13}=k_{31}$) are added in the model to account for a microwave field in resonance with the $\vert 1 \rangle \leftrightarrow \vert 2 \rangle$ (resp. $\vert 1 \rangle \leftrightarrow \vert 3 \rangle$) electron spin transition in the ground level. In the following ESR spectra are recorded with continuous-wave (CW) optical excitation. In this case, the mean PL rate is given by $\bar{\mathcal{R}}(\mathbf{B})=\eta \sum_{i=4}^6 \sum_{j=1}^3 \bar{n_i} k_{ij}$ where the averaged populations $\lbrace \bar{n_i} \rbrace$ are inferred by solving the classical rate equations (Eq.~(\ref{rate})) at the steady state. In this framework, the ESR contrast of the $\vert 1 \rangle \leftrightarrow \vert 2 \rangle$ transition is defined by    
\begin{equation}
\label{Contrast}
\mathcal{C}(\mathbf{B})=\frac{\bar{\mathcal{R}}(k_{12}=0)-\bar{\mathcal{R}}(k_{12})}{\bar{\mathcal{R}}(k_{12}=0)} ,
\end{equation} 
where $\bar{\mathcal{R}}(k_{12}=0)$ (resp. $\bar{\mathcal{R}}(k_{12})$) denotes the NV defect PL rate without applying the microwave field (resp. with a resonant microwave field). A similar expression holds for the $\vert 1 \rangle \leftrightarrow \vert 3 \rangle$ transition.

\subsection{Experimental setup}

Individual NV defects hosted in a high-purity type IIa diamond crystal (Element Six) are optically isolated at room temperature using a scanning confocal microscope. Optical excitation at $532$~nm wavelength is provided either by a CW laser or by a pulsed laser with a pulse duration $\delta t\approx 60$ ps and a $1/T=10$~MHz repetition rate (PicoQuant, LDH-P-FA-530B). The detection system comprises a confocal arrangement, an avalanche photodiode working in the single-photon counting regime (Perkin-Elmer, SPCM-AQR-14), and a time-correlated single-photon counting module with a bin size of $512$~ps (PicoQuant, PicoHarp 300). An electromagnet is used to apply a static magnetic field with controlled amplitude, while ESR transitions in the ground level are driven with a microwave field applied through a copper microwire directly spanned on the diamond surface. 
 
\subsection{Results and discussion}
\begin{figure}[t]
\begin{center}
   \includegraphics[width=0.92\textwidth]{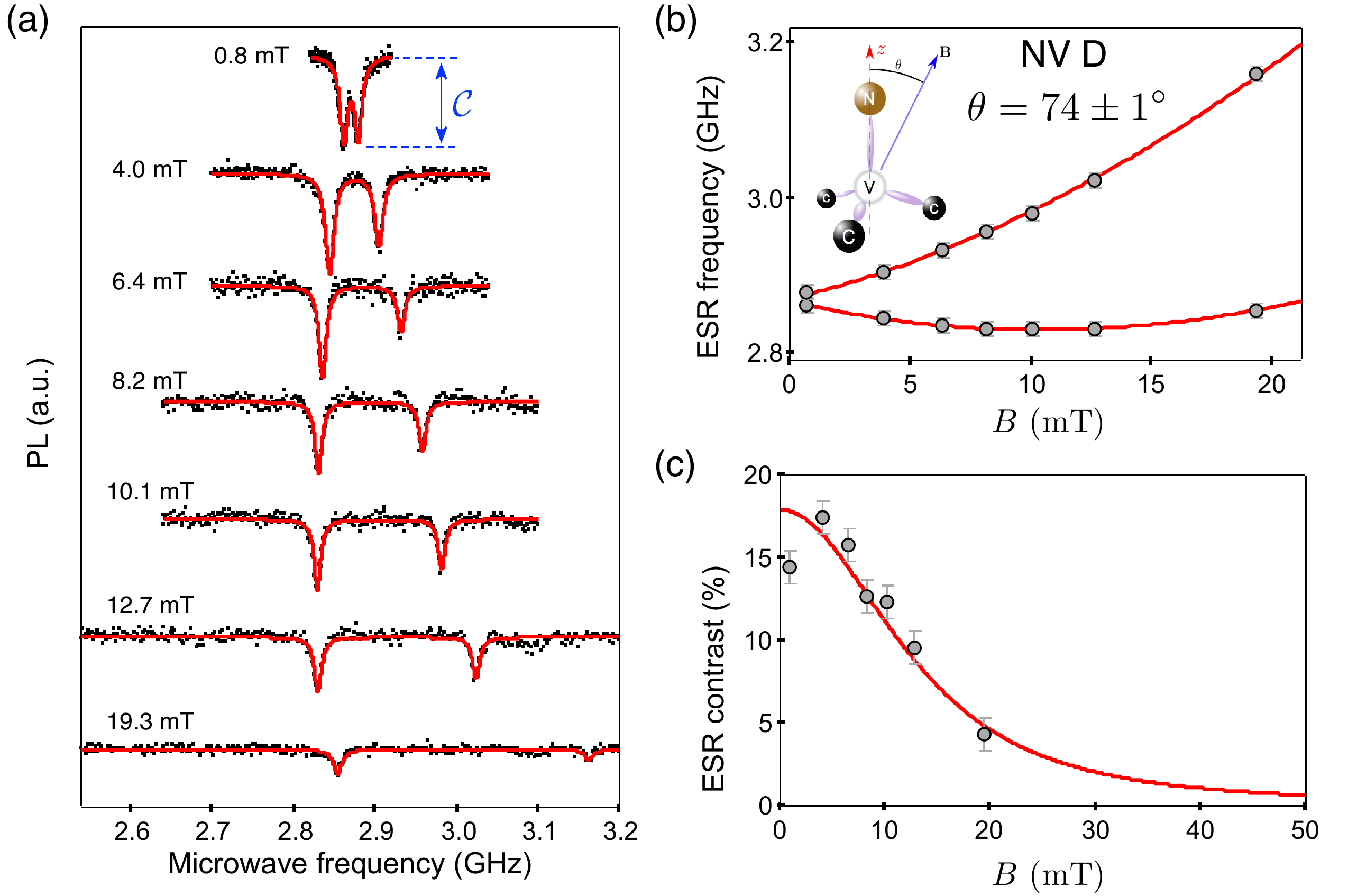}
\caption{(a)-Optically detected ESR spectra recorded for a single NV defect (NV~D) at various amplitudes $B$ of the magnetic field. Solid lines are data fitting with Lorentzian functions. (b)-ESR frequencies as a function of $B$. Solid lines are data fitting with the eigenenergies of ${\cal H}_{\rm gs}$. The angle deduced from the fit is $\theta=74\pm1^\circ$. (c)- ESR contrast as a function of $B$. Markers are data corresponding to the $\vert 1 \rangle \leftrightarrow \vert 2 \rangle$ electron spin transition while the solid line is a fit using Eq.~(\ref{Contrast}) of the seven-level model. Here the only free parameters are the optical pumping parameter $\beta$ and the ESR transition rate $k_{12}$. All other parameters of the model are obtained through time-resolved PL measurements as summarized in Table~1. For these experiments the NV defect was excited with a CW laser power of $100 \ \mu$W, leading to a detected PL intensity of $5\times 10^4$~counts.s$^{-1}$ at zero field. The evolution of the normalized PL intensity of NV~D as a function of the magnetic field strength is shown in Fig.~\ref{fig3}(c).}
\label{Fig0}
\end{center}
\end{figure}

\indent For each studied NV defect, the angle $\theta$ between its symmetry axis ($z$) and the magnetic field $\mathbf{B}$ was first measured by using ESR spectroscopy (see Tab.~\ref{tab1}). For that purpose, optically-detected ESR spectra were recorded by sweeping the frequency of a microwave field while monitoring the NV defect PL intensity. Owing to spin dependent PL, two dips can be observed in the PL signal, corresponding to electron spin transitions $\vert 1 \rangle \leftrightarrow \vert 2 \rangle$ and $\vert 1 \rangle \leftrightarrow \vert 3 \rangle$~(Fig. \ref{Fig0}(a)). Measuring the electron spin transition frequencies as a function of the magnetic field amplitude $B=\Vert{\bf B}\Vert$ allows us to extract the angle $\theta$ through data fitting with the eigenenergies of the ground level spin Hamiltonian ${\cal H}_{\rm gs}$~(Fig. \ref{Fig0}(b))~\cite{Balasubramanian2008}. 
\indent The ESR spectra depicted in Figure~\ref{Fig0}(a) can also be used to estimate the ESR contrast $\mathcal{C}$ as a function of the magnetic field amplitude (Fig.~\ref{Fig0}(c)). If the magnetic field is such that ${\cal H}_{\rm gs,es}^z\gg {\cal H}_{\rm gs,es}^\perp$ (see Eq.~(\ref{Hamilto})), the quantization axis is fixed by the NV defect axis, and a high ESR contrast is observed. Conversely, when the condition ${\cal H}_{\rm gs,es}^z\gg {\cal H}_{\rm gs,es}^\perp$ is not fulfilled, the quantization axis is rather determined by the applied magnetic field and $m_s$ is no longer a good quantum number. The eigenstates of the spin Hamiltonian are then given by superpositions of the $m_s=0$ and $m_s=\pm1$ spin sublevels, both in the ground and excited levels (see Eq.~(\ref{eigen})). As a result, optically-induced spin polarization and spin-dependent PL of the NV defect become inefficient, and the contrast of optically detected ESR vanishes, as shown in Figure~\ref{Fig0}(c). Magnetic field imaging through optically-detected ESR is therefore inefficient in the regime of `high' off-axis magnetic field. Nevertheless, the NV defect optical response can be used to extract information about the magnetic field in this regime, as explained below. \\
\begin{figure}[t]
\begin{center}
    \includegraphics[width=.92\textwidth]{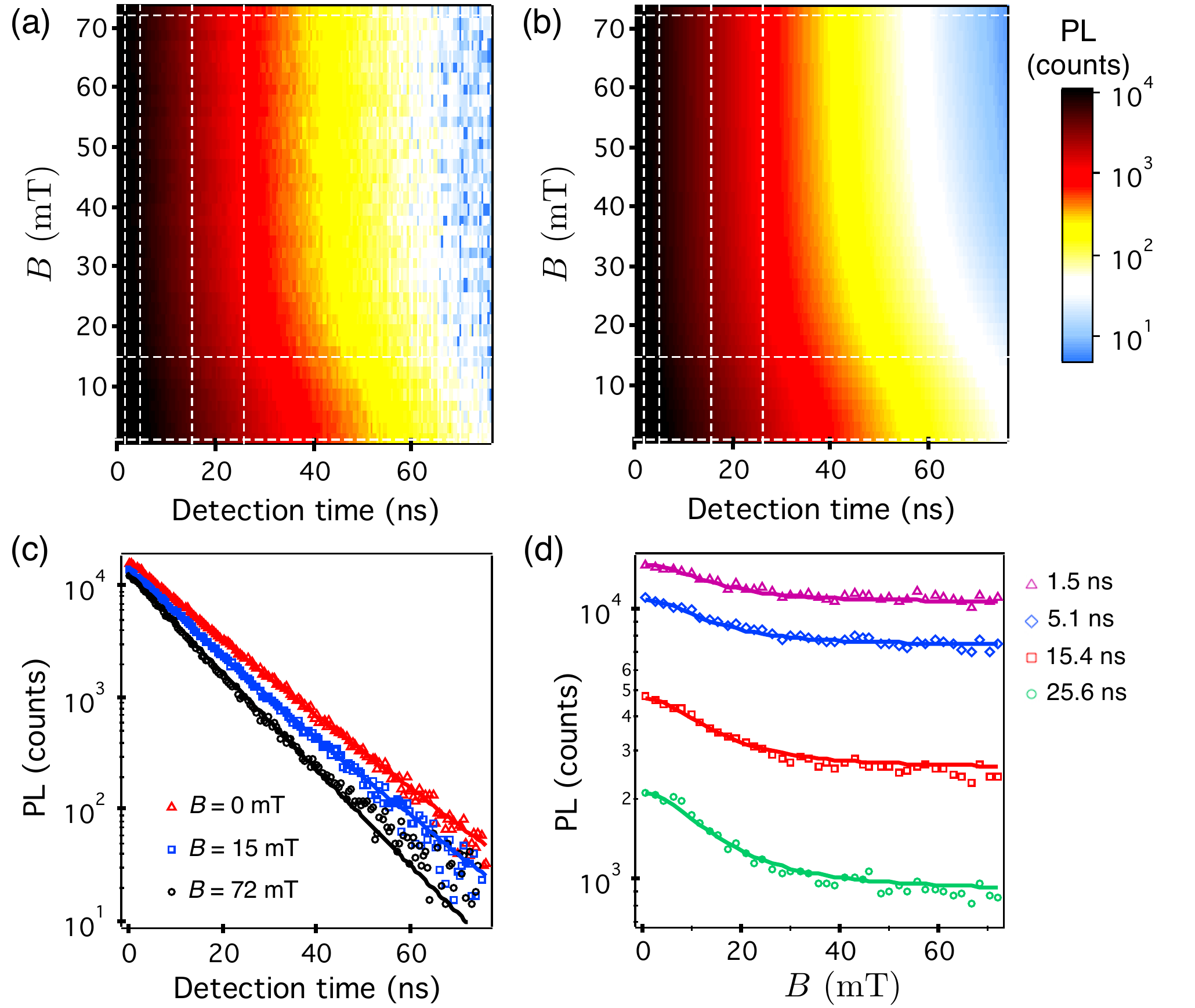}
\caption{(a)
      Time-resolved PL traces ${\cal R}(t,B)$ measured for NV E as a
      function of the magnetic field amplitude $B$. PL counts,
      corresponding to a 15 s integration time, are encoded in a
      logarithmic color scale. (b) Fit to the data as explained in the
      main text. The reduced chi-square error of the fit is $\chi^{2}_{\rm red}=1.15$. The parameters extracted from the fit are summarized in
      Table \ref{tab1}. (c) Linecuts corresponding to horizontal dashed
      lines in (a) and (b) showing several decay traces measured for
      various values of $B$ (markers) together with the fit (solid
      lines). At zero-field (red line), the relative amplitudes of the
      three exponential decays are 0.72, 0.14 and 0.14, corresponding to
      decay times $\tau_4=13.7$ ns, $\tau_5=8.6$ ns and $\tau_6=8.6$ ns,
      respectively. At the highest field (black line) they are
      found to be 0.34, 0.34 and 0.32, with decay times $\tau_4=11.0$
      ns, $\tau_5=9.4$ ns and $\tau_6=9.1$ ns, respectively. (d)
      Linecuts corresponding to vertical dashed lines in (a) and (b)
      showing measured PL counts as a function of $B$ at various
      detection times (markers) together with the fit (solid lines).}
\label{fig2}
\end{center}
\end{figure} 
      
\indent The photophysical parameters of the NV defect were first measured in order to compute its optical response as a function of the $\mathbf{B}$ field. For that purpose, time-resolved PL traces ${\cal R}(t,B)$ were recorded while increasing the magnetic field amplitude $B$ for a given angle $\theta$, as shown in Figure~\ref{fig2}(a). Fitting the whole set of data with Eq.~(\ref{eqn8}) allows us to extract the zero-field transition rates $\lbrace k_{r}^0,k_{47}^0,k_{57}^0,k_{71}^0,k_{72}^0\rbrace$ as well as the normalization coefficient $\eta$ and optical pumping parameter $\beta$~(Fig.~\ref{fig2}(b)). Horizontal and vertical linecuts of the matrix ${\cal R}(t,B)$ are plotted in Figure~\ref{fig2}(c) and (d), respectively, showing good agreement between the experimental data and the model. We emphasize the fact that such a tri-exponential fit can be reliably applied because all traces are fitted at once. We thus benefit from the correlations between traces at different magnetic field amplitudes, whereas each trace taken individually could not be fitted satisfactorily with a tri-exponential function. The transition rates extracted from the fitting procedure are summarized in Table~\ref{tab1} for four different NV defects oriented with different angles $\theta$ with respect to the magnetic field. We note that these values are in agreement with those reported in Ref. \cite{Robledo2011}, which were obtained using a different approach based on multi-pulse excitation techniques combined with coherent electron spin manipulation. As expected, we measure a spin-dependent ISC rate $k_{57}^0\gg k_{47}^0$, which is responsible for optically-induced spin polarization and spin-dependent PL response of the NV defect. Furthermore, our measurements seem to confirm that the metastable state decays roughly as often in the $m_s=0$ state as in the $m_s=\pm1$ state, {\it i.e.} $k_{71}^0\sim k_{72}^0$~\cite{Robledo2011}, in contradiction with the initial belief in a coupling to $m_s=0$ alone~\cite{Manson2006}. \\

\begin{table}[t]
\begin{center}
\begin{tabular}{ccccc}
\hline
& NV C & NV D & NV E & NV F \\
\hline
\hline
$\theta$ & $35 \pm 1^{\circ}$ & $74 \pm 1^{\circ}$ & $55 \pm 1^{\circ}$ & $36^{\circ} \pm 1^{\circ}$ \\
\hline
$k_r^0$ $(\mu$s$^{-1})$ & $67.7 \pm 3.4$ & $63.2 \pm 4.6$ & $63.7 \pm 4.5$ & $69.1 \pm 1.6$ \\
$k_{47}^0$ $(\mu$s$^{-1})$ & $6.4 \pm 2.3$ & $10.8 \pm 4.1$ & $9.3 \pm 3.0$ & $5.2 \pm 0.8$  \\
$k_{57}^0$ $(\mu$s$^{-1})$ & $50.7 \pm 4.4$ & $60.7 \pm 6.6$ & $53.0 \pm 5.9$ & $48.6 \pm 1.9$ \\
$k_{71}^0$ $(\mu$s$^{-1})$ & $0.7 \pm 0.5$ & $0.8 \pm 0.6$ & $0.9 \pm 0.8$ & $1.5 \pm 0.5$ \\
$k_{72}^0$ $(\mu$s$^{-1})$ & $0.6 \pm 0.3$ & $0.4 \pm 0.2$ & $0.5 \pm 0.2$ & $1.4 \pm 0.2$ \\
\hline
\end{tabular}
\caption{Summary of the parameters extracted from the fit of time-resolved PL traces ${\cal R}(t,B)$ for four different NV defects, oriented with different angles $\theta$ with respect to the magnetic field. The indicated values are mean $\pm$ one standard deviation, calculated from a set of fitting parameters obtained after running the fitting procedure many times while varying the initial guess.}
\label{tab1}
\end{center}
\end{table}

\begin{figure}[t]
\begin{center}
    \includegraphics[width=.93\textwidth]{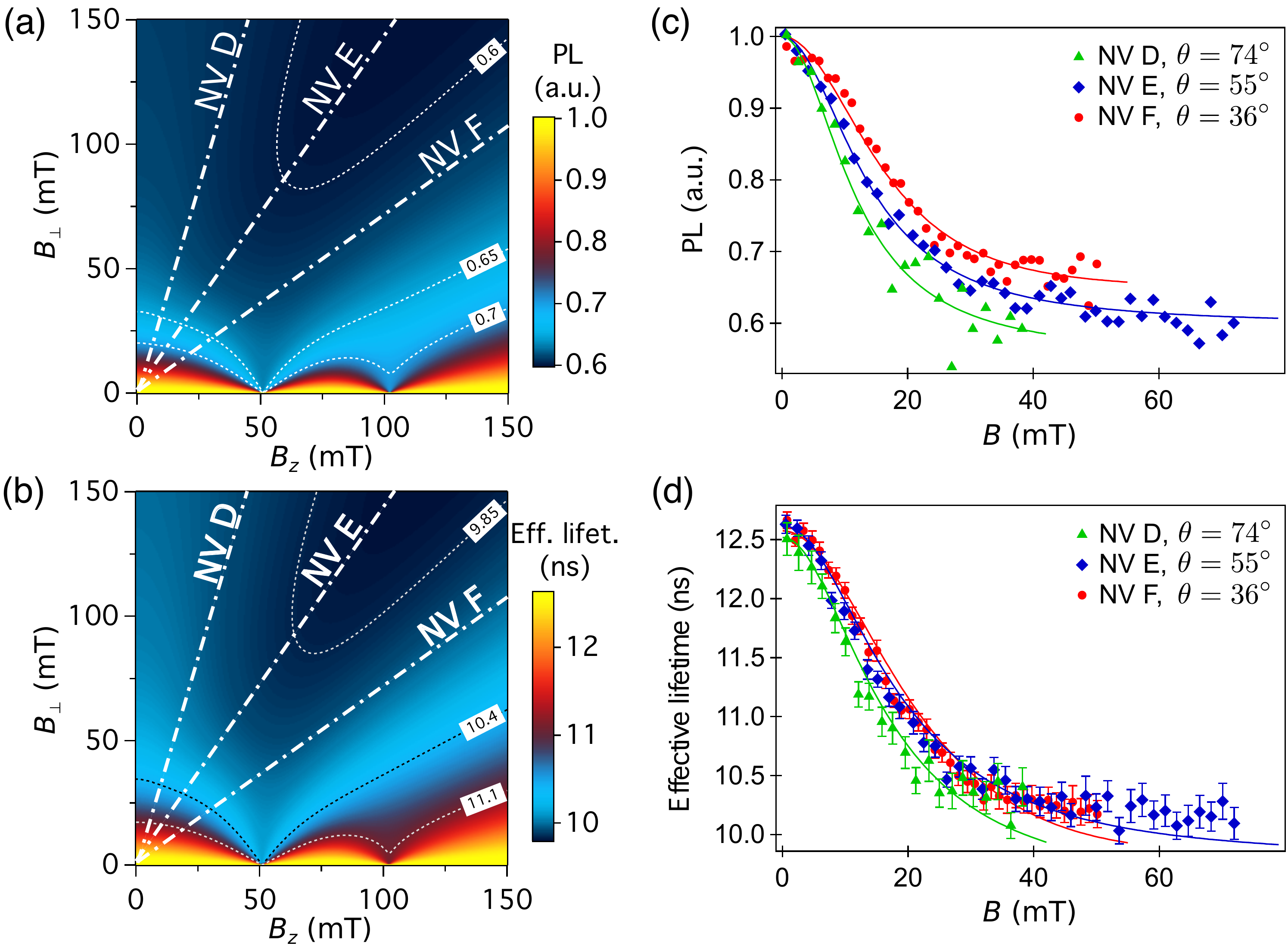}
\caption{(a) Calculated PL intensity and (b) effective lifetime $\tau_{\rm eff}$ as a function of the field components $B_z$ and $B_\perp$, parallel and perpendicular to the NV defect axis, respectively. For both (a) and (b), the photophysical parameters of NV E are used in the seven-level model, and a few contour lines are superimposed (dotted lines). (c) Integrated PL intensity and (d) effective lifetime as a function of $B$ measured for NV D, NV E and NV F. Error bars are 95\% confidence intervals. The angles $\theta$ corresponding to the three NVs considered in (c) and (d) are pictured by dashed lines in (a) and (b). Solid lines are the results of the calculation.}
\label{fig3}
\end{center}
\end{figure}

Having completely characterized the NV defect and extracted its photophysical parameters, we now discuss the dependence of the NV defect optical properties on the $\mathbf{B}$ field and its application for magnetic field imaging. Figure \ref{fig3}(a) shows the normalized PL intensity calculated using the seven-level model with the parameters of NV E  as a function of $B_z=B\cos\theta$ and $B_\perp=B\sin\theta$, varied in the range 0-150 mT. Regardless of the value of $B_z$ in this range, the PL steadily decreases with an increasing $B_\perp$, with a PL drop exceeding $30 \%$ for $B_\perp>20$ mT. This effect results from a mixing of electron spin states both in the ground and excited levels, which enhances the probability of non-radiative ISC to the metastable level and reduces the PL intensity. We note that in the low transverse field regime, the PL intensity also exhibits the well-known sharp drops at $B_z\approx51$ mT and at $B_z\approx102$ mT, corresponding to spin mixing induced by a level anti-crossing within the excited and ground level, respectively~\cite{Epstein2005,Rogers2009}. In Figure \ref{fig3}(c), the normalized integrated PL signal is plotted as a function of $B$ for three distinct NV defects, oriented with different angles $\theta$ with respect to {\bf B}. The data are well reproduced by the model (solid lines) without any free parameters. This graph highlights the fact that the monotonous PL decrease with $B$ is a general feature of NV defects, occurring whenever {\bf B} is not aligned with $z$ to better than $\approx 20^\circ$ (see Fig. \ref{fig3}(a)). Such a PL drop can be used to discriminate between low and high transverse magnetic field regions. Although not fully quantitative, a PL measurement sets a lower bound to the magnetic field amplitude. For instance one deduces from the contour lines in Fig. \ref{fig3}(a) that a PL drop of 30\% implies that the field has an amplitude $B>20$ mT. This property can be used to develop a microscope capable of qualitatively mapping large magnetic field regions at the nanoscale. We note that this method could be used with any solid-state emitter exhibiting spin-dependent PL~\cite{Koehl_Nature2011}, with a characteristic magnetic field range given by its zero-field splitting (Eq.~(\ref{Eq1})). \\

The PL drop induced by spin mixing is correlated with a reduction of the effective lifetime $\tau_{\rm eff}$ of the NV defect excited level (Fig.~\ref{fig2}(c)), which is defined as the exponential decay constant that best fit a given time-resolved PL trace. Using the photophysical parameters of NV E, we infer the effective lifetime $\tau_{\rm eff}$ as a function of $B_z$ and $B_\perp$ by fitting the time-resolved PL traces calculated from the seven-level model with a single exponential decay. As shown in Figure \ref{fig3}(b), the evolution of $\tau_{\rm eff}$ as a function of the magnetic field is clearly correlated with the PL rate (Fig. \ref{fig3}(a)). This feature illustrates that field-induced spin level mixing increases the mean probability for non radiative ISC transitions to the metastable level, which not only decreases the PL rate but also increases the decay rate from the excited level, thus leading to an overall reduction of the excited level lifetime. In Figure \ref{fig3}(d), $\tau_{\rm eff}$ is plotted as a function of $B$ for three different defects, together with the model that introduces no free parameters. In the same way as for the PL rate measurement, knowing the effective lifetime sets a lower bound for the amplitude of the magnetic field (see contour lines in Fig. \ref{fig3}(b)), and can be used to map high off-axis magnetic field regions, as demonstrated in the next section.

\section{All-optical magnetic field imaging with a scanning NV defect} \label{sec2}

\begin{figure}[t]
\begin{center}
    \includegraphics[width=1\textwidth]{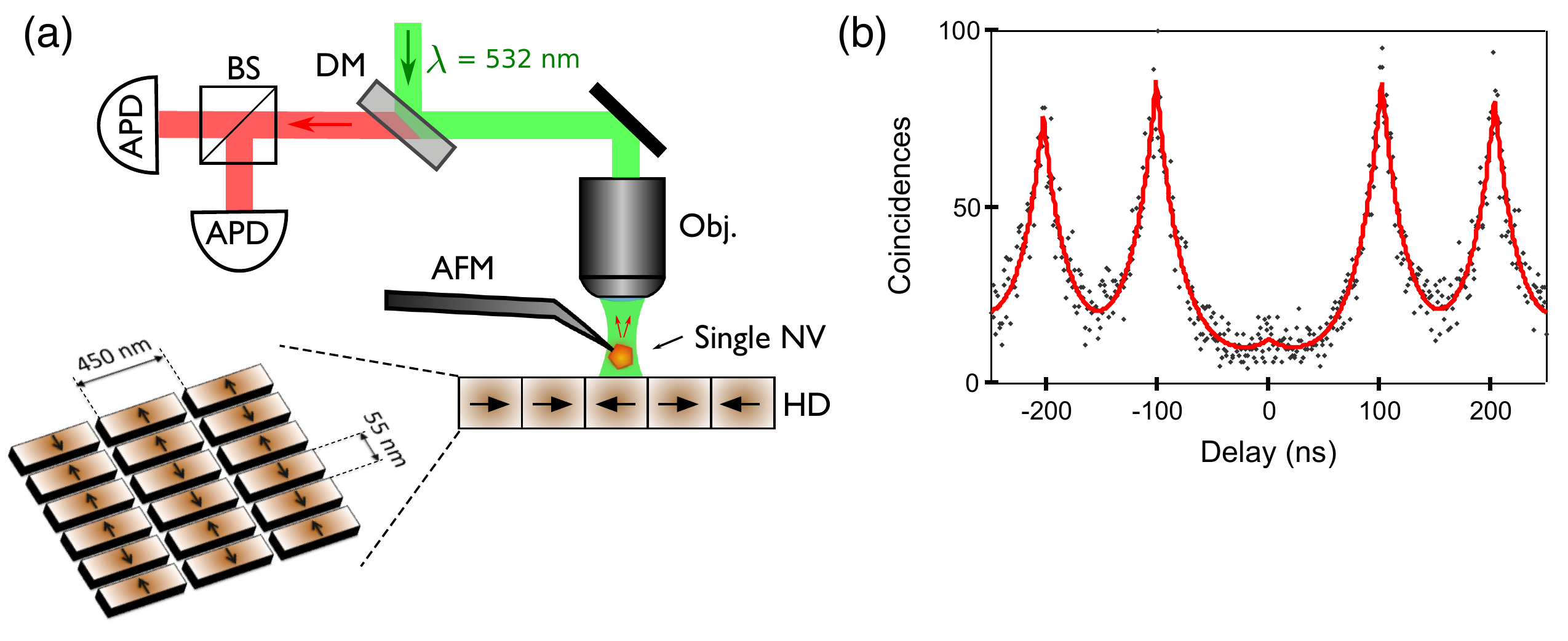}
\caption{(a) Simplified scheme of the scanning NV defect microscope. Either a pulsed or a CW laser source operating at $532$~nm wavelength is tightly focused at the end of the AFM tip through a microscope objective with a $0.9$-numerical aperture and a $1$-mm working distance (Olympus, MPLFLN100X). The PL emitted by a single NV defect placed at the apex of the tip is collected by the same objective and spectrally filtered from the remaining pump light using a dichroic mirror (DM). The collected light is then detected with a Hanbury Brown and Twiss (HBT) interferometer consisting in two single-photon counting modules (APD) placed on the output ports of a 50/50 beamsplitter (BS). Scanning probe magnetometry is performed by monitoring the magnetic-field-dependent NV defect PL while scanning the magnetic hard disk (HD). (b) Histogram of time delays between two consecutive single-photon detections, recorded in pulsed regime with the HBT detection system. The solid line is a data fit with multi-exponential functions.}
\label{fig4}
\end{center}
\end{figure}

Magnetic field imaging with a scanning NV defect is performed with the experimental setup depicted in Figure~\ref{fig4}(a), which combines an optical confocal microscope and a customized tuning-fork-based atomic force microscope (Attocube Systems, CFM/AFM), all operating under ambient conditions. A 20-nm diamond nanocrystal hosting a single NV defect is first grafted at the end of the AFM tip following the procedure described in Refs.\cite{Rondin2012,Cuche2009}. A confocal microscope placed on top of the tip allows us both to excite and collect the NV defect magnetic-field-dependent PL. The unicity of the emitter placed at the apex of the AFM tip was checked by measuring in pulsed regime the histogram of time delays between two consecutive single-photon detections with a Hanbury Brown and Twiss interferometer. As shown in Figure~\ref{fig4}(b), a strong reduction of coincidences is observed around zero delay, which is the signature that a single NV defect is attached to the tip~\cite{Lounis_Nature2000}.

\begin{figure}[t]
\begin{center}
    \includegraphics[width=0.7\textwidth]{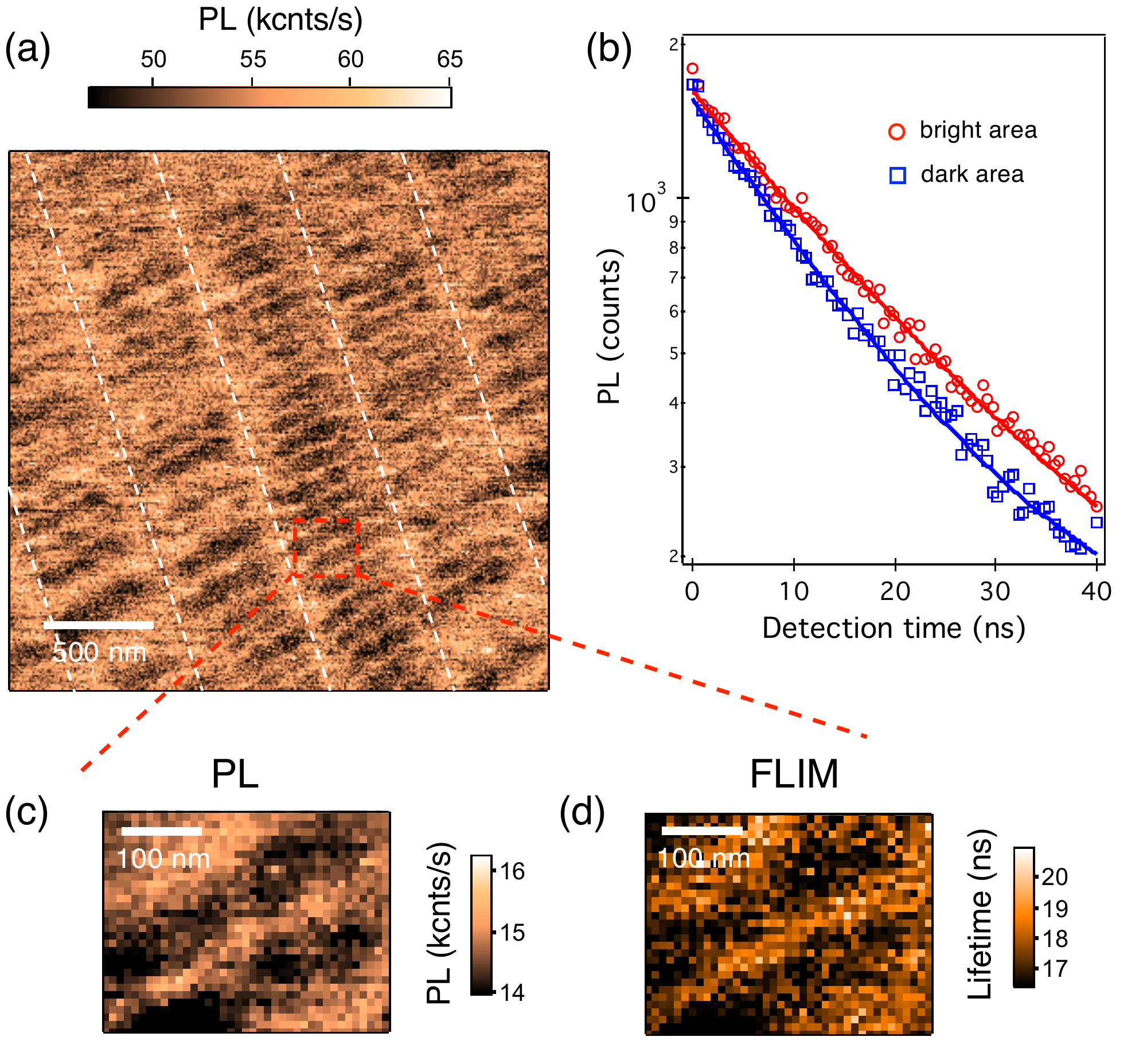}
\caption{(a) PL map recorded by scanning the magnetic hard disk while operating the AFM in tapping mode, with CW optical excitation. The image corresponds to $300\times 300$ pixels, with a $8$-nm pixel size and a $20$-ms acquisition time per pixel. The white dashed lines indicate the tracks of the magnetic hard disk. (b) PL decay traces recorded with the AFM tip standing either on a bright or on a dark area. Solid lines are data fits using single exponential decays. (c,d) PL (c) and effective lifetime (d) maps obtained under pulsed laser excitation. The scanned area (39 x 30 pixels) corresponds to the dashed red square depicted in (a) with a $3$-s acquisition time per pixel. }
\label{fig5}
\end{center}
\end{figure}

As a test sample, we used a piece of a commercial magnetic hard disk composed of $450$ nm x $55$ nm bits with random in-plane magnetization (Fig. \ref{fig4}(a)). Two adjacent magnetic bits with opposite magnetizations produce a stray field coming in or out at the boundary between the two bits. Imaging such a stray field is performed by recording the NV defect PL intensity while scanning the magnetic hard disk. In this experiment, the AFM is operated in tapping mode in order to maintain the mean probe-to-sample distance to a constant value. As shown in Figure~\ref{fig5}(a), the PL image reveals an array of dark areas that correspond to regions of high transverse magnetic field. Such dark areas indicate magnetization reversals, very much like in magnetic force microscopy (MFM) measurements with the difference that in addition MFM distinguishes between fields coming in or out of the sample, corresponding to repulsive or attractive forces. The distance between two consecutive dark lines is measured to be larger than 100 nm in Figure~\ref{fig5}(a). For instance, it is approximately $110$ nm in the zoom shown in Figure~\ref{fig5}(c), which corresponds to the size of two magnetic bits. We therefore believe that in this particular experiment, at least two consecutive bits with the same magnetization are needed in order to build up a stray field that is strong enough to induce a detectable PL quenching. 

As discussed in Section~\ref{sec1}, the field-induced PL quenching is associated with an overall reduction of the NV defect excited-level lifetime. This is illustrated in Figure~\ref{fig5}(b), which shows PL decay traces measured when the tip stands either on a bright area (low magnetic field) or on a dark area (high transverse magnetic field). Single exponential data fits give effective lifetimes $\tau_{bright}=18.2\pm0.3$ ns and $\tau_{dark}=15.1\pm0.2$ ns, corresponding to a reduction of $17\%$. We note that the lifetime of the NV defect is larger in nanocrystals compared to bulk measurements reported in Figure~\ref{fig3}(b)), owing to change of the surrounding refractive index~\cite{Beveratos_PRA2001}. By recording PL decay traces while scanning the sample, magnetic field mapping can also be achieved through fluorescence lifetime imaging (FLIM), as shown in Figure~\ref{fig5}(d)). As expected, FLIM and PL images are clearly correlated: the regions of low effective lifetime closely match those of low PL signal, in correspondence to regions of high off-axis magnetic field (Fig.~\ref{fig5}(c)-(d)).  

The maximum PL contrast in Fig. \ref{fig5}(b) is about 30\%, which is close to what was measured for NV defects in bulk diamond. As discussed in Section~\ref{sec1}, measuring the normalized PL intensity can in principle give access to a lower bound for $B$ provided that one knows all relevant parameters of the experiment, including the transition rates of the NV defect employed. However, those are affected by the electromagnetic environment of the NV defect -- proximity to the AFM tip, to the metallic sample, etc. -- and are therefore not readily accessible. Assuming that the general behavior of Figure~\ref{fig3}(a) is however still approximately valid here, one can state that the dark regions in Figure~\ref{fig5}(a) correspond to a magnetic field exceeding $10$ mT. This would be consistent with the stray field expected for such a magnetic hard disk at a distance of a few tens of nanometers~\cite{Rondin2012}. Although not quantitative, all optical magnetic field imaging with a scanning NV defect is thus a relatively simple way to map regions of magnetic field larger than a few tens of mT at the nanoscale. Furthermore, unlike MFM whose magnetic tip is likely to perturb the magnetization of the studied sample, the NV defect probe introduces no magnetic back action to the sample at first order.    

Finally, we note that the ability to perform fluorescence lifetime imaging (FLIM) with a scanning NV defect opens up a large range of possibilities in the field of nanophotonics. Indeed, mapping the local density of electromagnetic states (LDOS) above non-magnetic samples is in reach, with the potential for vectorial LDOS mapping if one uses NV defects with different orientations. Compared to recent demonstrations of scanning FLIM~\cite{Farahani2005,Frimmer2011}, advantages of NV defects are their perfect photostability as well as their atomic-sized spatial resolution. In this context, this work is a starting-point for understanding and quantifying the relation between the NV defect dynamics and the LDOS, a necessary step for future LDOS mapping experiments.      

\section{Conclusion} \label{sec3}

In this article the optical properties of NV defects in diamond have been studied as a function of the external magnetic field. We performed time-resolved measurements and developed a seven-level model that accounts for the decreased ESR contrast, PL intensity and effective lifetime when a transverse magnetic field is applied. We demonstrated an application of those effects to nanoscale magnetic imaging. Using a scanning NV defect microscope, we mapped the stray field of a magnetic hard disk by recording either PL or lifetime images. This all-optical method for high magnetic field imaging might be of interest in the field of nanomagnetism, where samples producing fields in excess of several tens of milliteslas are typical.


\section*{Acknowledgments}

The authors acknowledge A. Dr{\'e}au, L. Mayer, F. Grosshans, S. Rohart, and A. Thiaville for fruitful discussions. This work was supported by the Agence Nationale de la Recherche (ANR) through the projects D{\sc iamag} and A{\sc dvice}, by C'Nano Ile-de-France and by RTRA-Triangle de la Physique (contract 2008-057T).

\end{document}